\documentclass[useAMS,usenatbib,fleqn]{mn2e}

\usepackage{txfonts}
\usepackage[pdftex]{graphicx}
\usepackage{graphicx}
\usepackage{natbib}
\usepackage{subfigure}

\bibpunct{(}{)}{;}{a}{}{,}

\usepackage{times}

\newcommand{\apj}{ApJ}
\newcommand{\apjl}{ApJL}
\newcommand{\apjs}{ApJS}
\newcommand{\aj}{AJ}
\newcommand{\mnras}{MNRAS}
\newcommand{\nat}{Nature}

\newcommand{\aap}{A\&A}
\newcommand{\apss}{Ap\&SS}
\newcommand{\aaps}{A\&AS}

\newcommand{\physrep}{PhysRep}

\title{Effects of superstructure environment on galaxy groups}

\author[Luparello et al.]
   {H.~E.~Luparello$^{1}$, M.~Lares$^{1}$, C.Y.~Yaryura$^{1}$, D.~Paz$^{1}$,
   N.~Padilla$^{2,3}$ and  D.~G.~Lambas$^{1}$\\
 $^{1}$Instituto de Astronom\'{\i}a Te\'{o}rica y Experimental
  (CONICET-UNC). Observatorio Astron\'{o}mico de C\'{o}rdoba, 
  Laprida 854, X5000BGR, C\'{o}rdoba, Argentina\\
 $^{2}$Departamento de Astronom\'{\i}a y Astrof\'{\i}sica, Pontificia
 Universidad Cat\'olica de Chile, Santiago, Chile\\
 $^{3}$Centro de Astro-Ingenier\'{i}a, Pontificia Universidad Cat\'olica de Chile, Santiago, Chile}

\date{Released 2012 Xxxxx XX}
\pagerange{\pageref{firstpage}--\pageref{lastpage}} \pubyear{2012}

\def\LaTeX{L\kern-.36em\raise.3ex\hbox{a}\kern-.15em
    T\kern-.1667em\lower.7ex\hbox{E}\kern-.125emX}

\begin{document}
\label{firstpage}
\maketitle
\begin{abstract}
We analyse properties of galaxy groups and their dependence on 
the large-scale environment as defined by superstructures.
We find that group--galaxy cross--correlations depend only on group 
properties regardless the groups reside in superstructures.
This indicates that the total galaxy density profile around groups 
is independent of the global environment.
At a given global luminosity, a proxy to group total mass, groups 
have a larger stellar mass content by a factor 1.3, a relative 
excess independent of the group luminosity. 
Groups in superstructures have 40 per cent higher velocity dispersions 
 and systematically larger minimal enclosing radii.
We also find that the stellar population of galaxies in groups in 
superstructures is systematically older as infered from the galaxy 
spectra $Dn_{4000}$ parameter.
Although the galaxy number density profile of groups is independent 
of environment, the star--formation rate and stellar mass profile of the 
groups residing in superstructures differs from groups elsewhere.
For groups residing in superstructures, the combination 
of a larger stellar mass content and star--formation rate produces a larger 
time--scale for star formation regardless the distance to the group center. 
Our results provide evidence that groups in superstructures formed 
earlier than elsewhere, as expected in the assembly bias scenario.
\end{abstract}


%····································································

%····································································

\section{Introduction} \label{S_intro}
%{{{S_intro*/
The large--scale structure of the Universe appears as a network 
made up of walls, filaments, knots and voids \citep{Joeveer:1978, 
Gregory:1978, Zeldovich:1982, deLapparent:1986}.
The nodes are the intersections of walls and filaments, so they 
are the highest density regions, usually known as superclusters.
Supercluster of galaxies are the largest systems present in the Universe.
There are many galaxy groups and clusters inhabiting these 
large systems.
Studies of the galaxy clusters according to their environment 
are useful to understand the properties and evolution of both, 
galaxy clusters themselves and the large--scale structure of the Universe.
Besides, studying their influence on galaxy clusters it is possible 
to analyse properties and the evolution of large--scale structure.
Recent galaxy redshift surveys (e.g. \citet{York:2000, Colless:2001}) sample
a sufficiently large volume to allow the study of the influence of 
supercluster in galaxy groups.
There are many previous studies, both theoretical and observational, 
that show that richer and more luminous groups and 
clusters of galaxies are located in higher density environments 
\citep{Einasto:2003b,Einasto:2005b,Croft:2011}; while the galaxy
properties, such as the star--formation rate or galaxy colours, 
also depend on large--scale structure \citep{Binggeli:1982, Donoso:2006, 
Crain:2009, White:2010}.
\citet{Einasto:2007c} analised the properties of galaxies 
in superclusters in the 2dFRS galaxy catalogue, and found that galaxy 
morphologies and their star formation activity are influenced by both 
the local and global environments.
\cite{Park:2009} used a volume--limited sample of galaxies extracted 
from the SDSS--DR4 to show that the influence of the large--scale 
density is not very significant over several galaxy properties once 
luminosity and morphology are fixed. 
However they suggest that this weak residual effect is due to 
the dependence of halo gas property on the large--scale density. 
Regarding to numerical simulations, \citet{Gil:2011} took into account 
the environmental influence proposing an extension of the halo model.
The formation and evolution of systems that are embedded in 
superstructures could be conditioned by these large overdensities 
\citep[][ and references therein]{Hoffman:2007, Araya-Melo:2009, 
Bond:2010, Pompei:2012}.
In high density regions clusters have a larger amount of substructures 
and higher peculiar velocities of their main galaxies than in low density 
regions \citep{Einasto:2005b,Tempel:2009}.

Most recent results are presented in \citet{Einasto:2012,Lietzen:2012} and
\citet{CostaDuarte:2012}.
Distinguishing between spider and filamentary morphology of superclusters, 
\citet{Einasto:2012} found that clusters in spider shape superclusters 
tend to have more substructure and higher peculiar velocities than cluster 
in filamentary superclusters.
Furthermore, clusters that are not members of superclusters have less 
substructure and lower values of peculiar velocities than supercluster 
members \citep{Einasto:2012}.
\citet{CostaDuarte:2012} verified the results found by 
Einasto et al. (2012) and also studied the effect of environment 
on galaxies in clusters and their outskirts. 
They suggest that the stellar population of clusters does not 
depend on supercluster richness nor morphology.
\citet{Lietzen:2012} studied how the galaxy evolution is influenced by the 
local group scale and the large scale environment. 
They found that in voids, the fraction of passive and star--forming galaxies 
in groups are approximately equal, while in superclusters the fraction of 
passive galaxies are considerably larger than those of star--forming galaxies.
Moreover, equally rich groups are more luminous in superclusters than in voids.
In this work, we will analyse the properties of galaxy groups 
located in large superclusters.

%%%%%%%%%% This paper is organized as follows:
 
This paper is organized as follows.
In Section 2 we present the data samples extracted from the galaxy, group and
supercluster catalogues. 
In Section 3 we show the results of the analysis of
group properties and their dependence with the large scale environment
characterized by superstructures.
Discussion and conclusions and given in Section 4.
Throughout this paper, we adopt a concordance cosmological model
($\Omega_{\Lambda}=0.75$, $\Omega_{matter}=0.25$) in the calculation
of distances.

%}}}S_intro*/

\section{Data and Samples} \label{S_data}
%{{{S_data*/

\subsection{SDSS--DR7} \label{SSDS-DR7} 

One of the largest and most ambitious surveys carried out so far 
is the Sloan Digital Sky Survey \citep{York:2000}. 
It has deep multi-color images covering more than one quarter 
of the sky and creates three-dimensional maps containing 
more than $930000$ galaxies and more than $120000$ quasars.  
Its main goal is the study of large-scale structure
of the Universe, producing also astronomical data for other areas. 
In particular, in this work we use the Seventh Data Release (DR7) of the
spectroscopic galaxy catalogue. 
It is one of the largest data sets produced by this project and contains images, 
image catalogues, spectra and redshifts.
The limiting magnitude of the spectroscopic galaxy catalogue in the r--band is $m_r=17.77$ 
\citep{Strauss:2002}.
The information about this survey is publicly available \footnote{http://www.sdss.org/dr7}.
We also use the MPA-JHU DR7 \citep{Kauffmann:2003}, which provides 
additional information about star--formation rates, stellar masses 
and the spectral indicator $Dn_{4000}$.
The star--formation rates (SFRs) are computed following the procedure 
described by \citet{Brinchmann:2004}.

Regarding to the stellar masses, they are obtained as explained in 
\citet{Kauffmann:2003} and \citet{Salim:2007}.
Spectral indicators are usually computed to quantify spectral
evolution of galaxies.
A widely used parameter is the $Dn_{4000}$ index, introduced by 
\citet{Hamilton:1985}.
The $4000 \AA$ break amplitude is obtained from the ratio of the
averaged flux density above $4000 \AA$ to the averaged flux density
below $4000 \AA$ \citep{Hamilton:1985}.
The ranges used to compute that averages are $4050-4250 \AA$ and
$3750-3950 \AA$.
The break at $4000 \AA$ is produced by 
the absorption of metalic lines, specially Fraunhofer H
and K lines of CaII, and lines of various elements heavier than
Helium in several stages of ionization.
The opacity for photons bluer than $\lambda \sim 4000 \AA$ rapidly
increases, producing a characteristic change in the intensity.
This intensity drop is enhanced in galaxies with metal rich, old
stellar populations.
Also, the Balmer lines close to $4000 \AA$ become broader and deeper
with time from the starburst \citep{Sanchez:2012}.
This parameter is a good indicator of stellar evolution, since it
correlates with effective temperature, surface gravity and metalicity
in stellar spectra \citep{Gorgas:1998, Gorgas:1999}.

\subsection{Catalogue of Superstructures} \label{superstructures}

The clustering properties of galaxies in scales smaller than the size
of superstructures are key to observationally constrain the accretion
process that give rise to luminous galaxies.
The evolution of the supercluster--void network depends on the matter/energy
contents of the universe, and thus are sensitive to the cosmological model.
Using numerical simulations of a LCDM universe evolved up to a scale factor 100
times the present value, \citet{Dunner:2006} find that there is a minimum mass
overdensity for a structure to remain bound in the future.
According to this, they stablish a criteria to identify structures in the
present day universe that are likely to evolve into virialized structures.
On the other hand, \citet{Einasto:2007} proposed a method to identify superstructures from a
smoothed luminosity density field.  
The authors build a superstructure catalogue by isolating large regions in
space that have a luminosity above a given threshold, calibrated so that the
largest superstructure is limited to the size of the largests known
superclusters.
The density field method can be combined with the results from numerical
simulations to establish, assuming a constant mass to luminosity relation,
criteria to isolate structures that are likely to evolve into virialized
structures in the distant future \cite{Luparello:2011}.
We use this catalogue of ``superstructures'', i.e., 
systems that are Future Virialized
Structures (hereafter FVS) to characterize the large scale
 environment of galaxies.
The catalogue of FVS was extracted from a volume-limited sample of galaxies
from the SDSS--DR7, with a limiting absolute magnitude of \mbox{$M_r <-20.47$}, 
in the redshift range \mbox{$0.04<z<0.12$}. 
The luminosity-density field is constructed on \mbox{1 h$^{-1}$ Mpc} cubic
cells grid, applying an Epanechnikov kernel of \mbox{ $r_0$ = 8 h$^{-1}$ Mpc}
(equation 3 of \cite{Luparello:2011}). 
The structures are constructed by linking overdense cells with
 a ``Friends of Friends'' algorithm, using a luminosity overdensity threshold of 
\mbox{$D_T=\rho_{lum}/\bar{\rho}_{lum}=5.5$}.
They also assign a lower limit for the total luminosity of a structure at \mbox{L$_{struct}\geq$
10$^{12}$L$_{\odot}$} to avoid contamination from smaller systems. 
The main catalogue of superstructures has completeness over 90 per cent and
contamination below 5 per cent, according to calibrations made using mock
catalogues.
The volume covered by the catalogue is \mbox{3.17 $\times$ $10^7$ (h$^{-1}$
Mpc)$^3$}, within which 150 superstructures were identified, composed by a
total of 11394 galaxies.
FVS luminosites vary between \mbox{ $10^{12}$ L$_{\odot}$} and \mbox{$\simeq$
10$^{14}$ L$_{\odot}$}, and their volumes range between \mbox{10$^2$ (h$^{-1}$
Mpc)$^3$} and  \mbox{10$^5$ (h$^{-1}$ Mpc)$^3$}.

The authors analysed 3 samples of SDSS--DR7 galaxies with different 
luminosity thresholds, dubbed S1, S2 and S3, and are described in 
Table 1 of their paper.
We will consider sample S2 in our analysis, which contains 89513
galaxies with $Mr <-20.47$ in the intermediate redshift range $0.04 <z
<0.12$.

\subsection{Galaxy Group Samples}

\begin{figure} 
   \centering
   \includegraphics[width=0.45\textwidth]{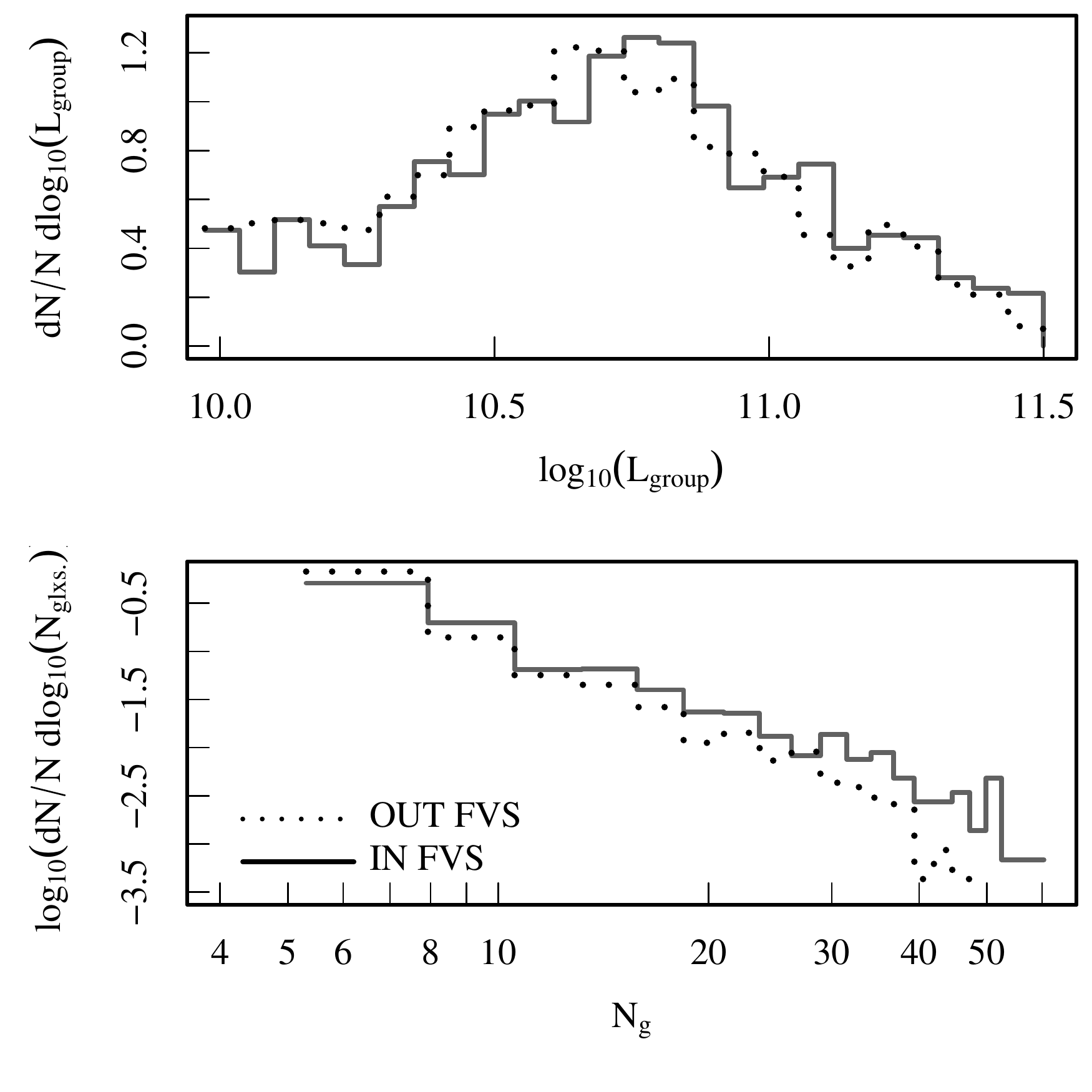}
   \caption{Luminosity (upper panel) and Multiplicity (bottom panel) 
distributions for the group samples inside (solid line) and outside (dotted line) superstructures.
        %Arrow indicates the higher luminosity threshold to define the samples.
   }
   \label{F1}
\end{figure}

The galaxy groups used in this work are identified in the SDSS-DR7
galaxy catalogue following \citet{Zapata:2009}.
The identification of these groups uses the same method presented in
\citet{Merchan&Zandivarez:2005}, who implemented a Friends of Friends
(FoF) algorithm, with a variable projected linking length $\sigma$,
with \mbox{$\sigma_0 = 0.239 h^{-1}Mpc$} and a fixed radial linking
length \mbox{$\Delta v = 450kms^{-1}$}. 
These values were set by \citet{Merchan&Zandivarez:2005} to obtain a
sample as complete as possible and with low contamination ($95 \%$ and
$8 \%$, respectively).
The variable linking lenght is calibrated to compensate the sample
dilution with redshift, as the original sample of galaxies is
flux-limited.
The full catalogue comprises 17.106 groups, in the redshift
range $0.001<z<0.5$.
With the aim to study the influence of superstructures in the global
properties of galaxy groups, we consider two subsamples of groups:
those residing in FVS and those which are not members of FVS.
It is well known that the luminosity correlates with the clustering
amplitude \citep{Alimi_1988, Zehavi:2005, Swanson_2008, Wang:2011,
Zehavi:2010, Ross_2011}.
In order to avoid possible effects of total group luminosity (a
suitable proxy to group total mass) we define samples in FVS and
elsewhere, $G_{in}$ and $G_{out}$ respectively, by requiring them to
have similar luminosity distributions.
With this restriction, our group samples are suitable to study
different properties of systems and their relation to environment.
The resulting samples $G_{in}$ and $G_{out}$, in the reshift range
$0.06 < z < 0.12$, contain 1457 and 2645 groups, respectively, with
the same luminosity distributions.
All these groups comprise at least 4 galaxies brighter
than $Mr <-20.47$, so that the final sample of galaxies in groups is
volume limited up to $z=0.12$.
In the upper panel of figure \ref{F1} we show the luminosity
distributions of samples $G_{in}$ and $G_{out}$ where it can be seen
their similarity in the total luminosity range
$10^{10}-10^{11.5}L_{\odot}$.

\begin{figure} 
   \centering
   \includegraphics[width=0.45\textwidth]{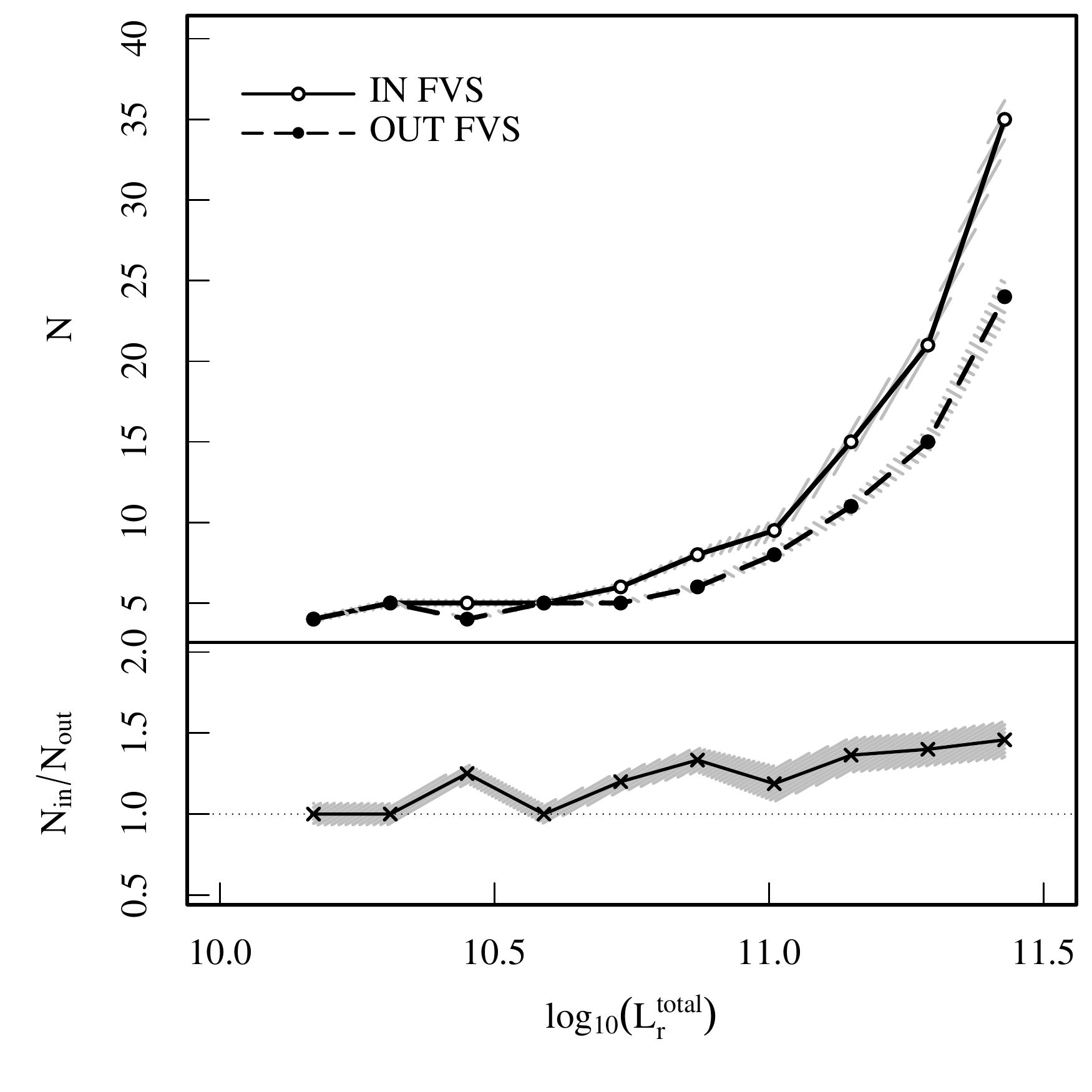}
   \caption{Upper panel: Mean multiplicity of groups, in bins of group mean
       luminosities, for groups in $G_{in}$ (solid line) and $G_{out}$ (dashed line).
        The error bands correspond to the error of the means.
        Bottom panel: ratio of multiplicity for group samples 
       $G_{in}$ and $G_{out}$.
      %Arrow indicates the higher luminosity threshold to define the samples.
   }
   \label{F0}
\end{figure}

Some studies have shown that properties of the galaxy groups correlate
with the group multiplicity.
\citet{Lietzen:2012} found that the dependence of the fraction of
different types of galaxies in groups varies with the richness of the
group.
These authors show that the fraction of star--forming galaxies
declines and the fraction of passive galaxies increases as the
richness of a group rises from one to approximately ten galaxies. 
They also show that the fraction of galaxies of different types do not
depend on group richness, for groups comprising between 20 and 50
members.
The lower panel of figure \ref{F1} shows the multiplicity
distributions of the samples $G_{in}$ and $G_{out}$.
In order to analyse multiplicity dependence on system luminosity, we
have computed the mean multiplicity in bins of total group luminosity.
The results are shown in figure \ref{F0} where it can be seen that the
number of galaxies in groups at a given total luminosity is larger in
FVS by a factor 1.3.
%%%%%%%%%%%%%%%%%%%%%%%%%%% R E S U L T S 

\begin{figure} 
   \centering
   \includegraphics[width=0.45\textwidth]{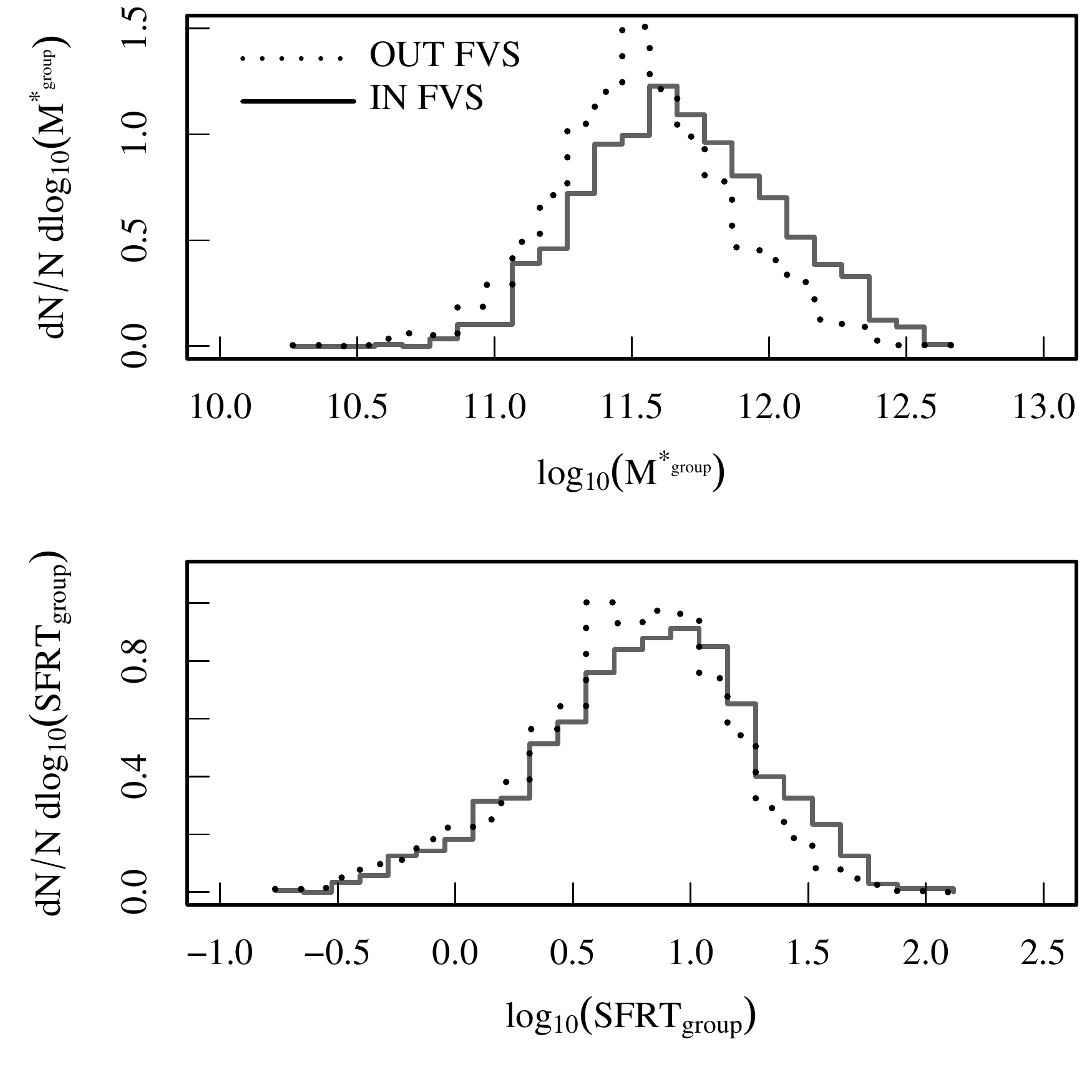}
   \caption{Stellar mass (upper panel) and total star--formation rate (bottom panel) 
distributions for the group samples inside (solid line) and outside (dotted line) superstructures.
   }
   \label{F2}
\end{figure}

\section{Properties of Galaxy Groups}

\subsection{Groups stellar mass, velocity dispersion and star formation time--scale}

\begin{figure}
 \centering
   \includegraphics[width=0.45\textwidth]{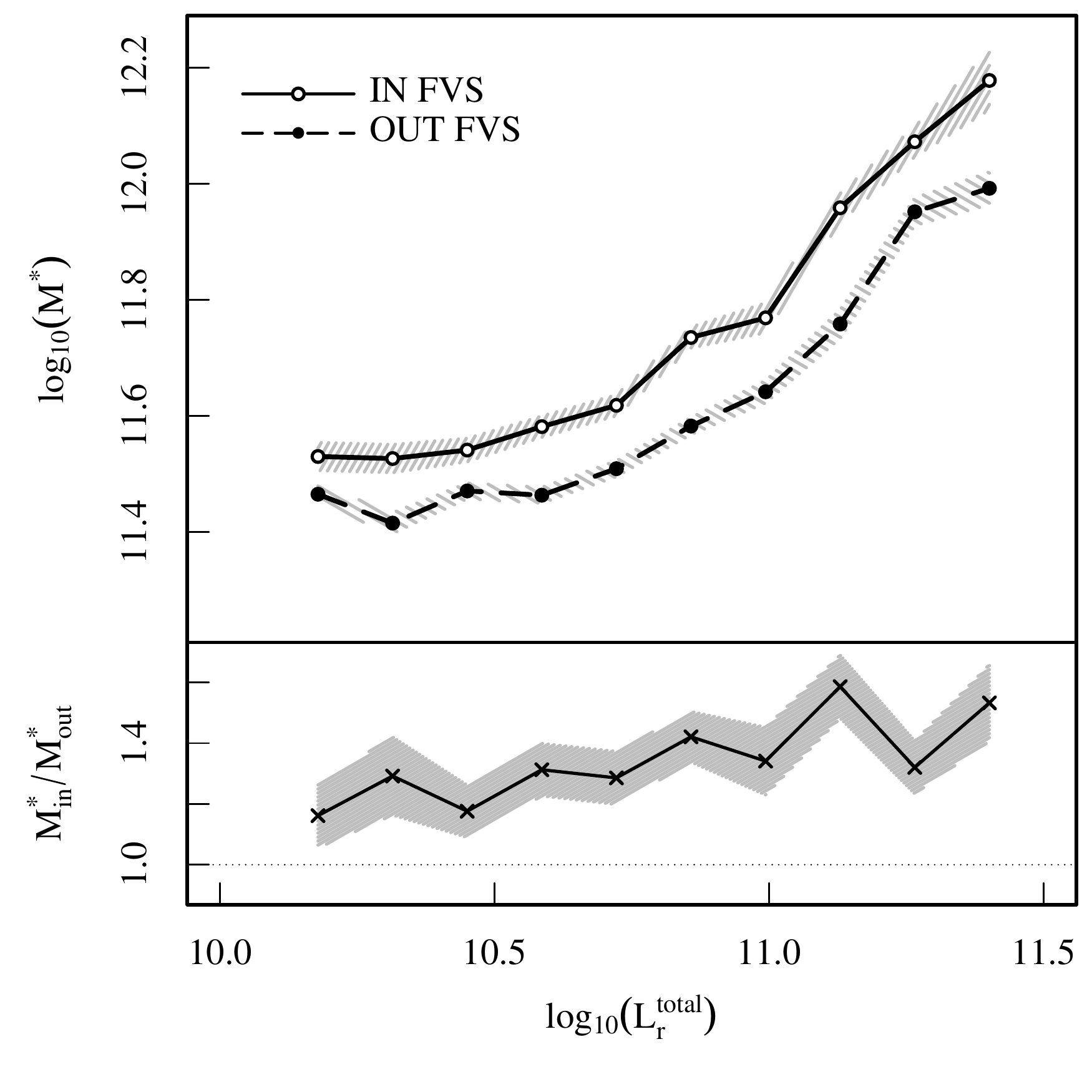}
   \caption{Upper panel: Mean total stellar mass of groups, in bins of group mean
       luminosities, for groups in $G_{in}$ (solid line) and $G_{out}$ (dashed line).
        The error bands correspond to the error of the means.
        Bottom panel: ratio of mean total stellar mass for group samples 
       $G_{in}$ and $G_{out}$.
   }
   \label{F_4}
\end{figure}

We have estimated the total stellar mass content of each group by
adding the stellar masses of their members. 
In a similar fashion, we have also computed the total star--formation
rate of the groups by adding the star--formation rates of the
individual members.
In figure \ref{F2} we show the resulting total group stellar mass (M$^{*}_{group}$) and
total group star--formation rate (SFRT$_{group}$) distributions for the two samples
$G_{in}$, and $G_{out}$. 
It can be seen that the total stellar mass is systematically 
larger for groups residing in superstructures. 
Also, the total star--formation rate distributions shows this same tendency.
The median total stellar mass content of groups in sample
$G_{in}$ is $4.6\times10^{11}M_{\odot}$, while the corresponding
median in sample $G_{out}$ is $3.4\times10^{11}M_{\odot}$. 
The star--formation rate medians are $6.7M_{\odot} yr^{-1}$ and $5.7 M_{\odot} yr^{-1}$ for
$G_{in}$ and $G_{out}$ group samples, respectively.
We have also explored the difference of total stellar masses as a function of group luminosity. 
In Fig. \ref{F_4} we show the mean total stellar mass as a function of 
group luminosity for $G_{in}$ and $G_{out}$ samples.
By inspection to this figure, and in particular to the lower panel, it can 
be seen that the excess of the total stellar mass for groups residing 
in FVS is not strongly dependent on group luminosity.
It should be noticed that the difference of the star--formation rate distributions
is less significant than the difference of the stellar mass distributions (17 per cent
excess in SFRT compared to 35 per cent in M* for sample $G_{in}$).
These results have a natural explanation in a scenario where groups in FVS started their
star--formation process earlier. 

\begin{figure} 
   \centering
   \includegraphics[width=0.45\textwidth]{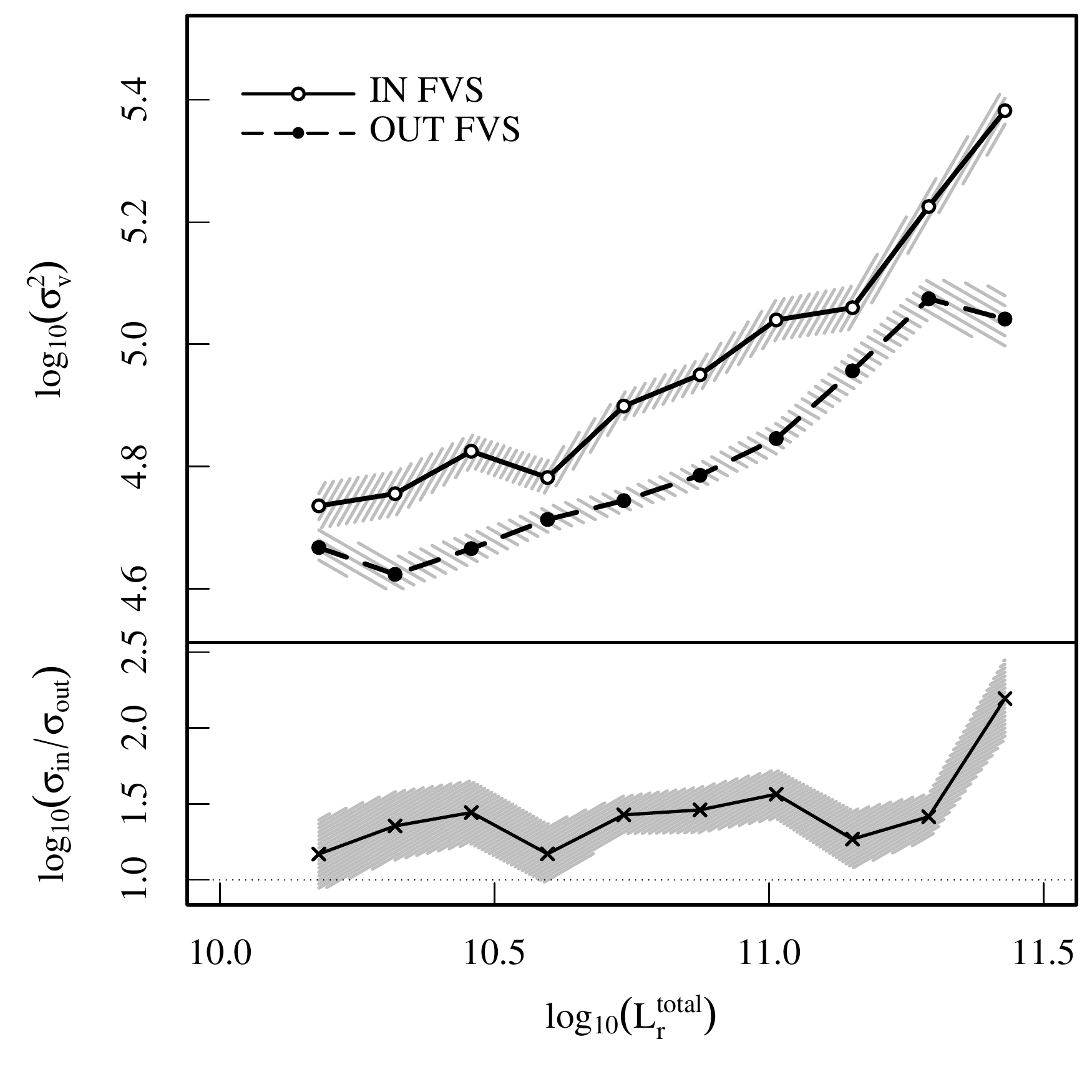}
   \caption{Upper panel: Mean velocity dispersions of groups, in bins of group mean
            luminosities, for groups in $G_{in}$ (solid line) and $G_{out}$ (dashed line).
            The error bands correspond to the error of the means. Bottom panel: ratio of mean
             velocity dispersions for group samples 
            $G_{in}$ and $G_{out}$.
   }
   \label{F3}
\end{figure}

Velocity dispersion of groups can be used to explore the dynamics of 
these systems. 
We have also studied the dependence of the groups velocity dispersion 
on group luminosity for samples $G_{in}$ and $G_{out}$.
The results are given in Fig. \ref{F3}, where it can be seen that groups 
located in FVS (sample $G_{in}$) present larger velocity dispersion than 
groups located in less dense environments, regardless of group luminosity.
This results is consistent with \citet{Einasto:2012} although we stress 
the fact that our samples $G_{in}$ and $G_{out}$ were selected to have 
groups with similar luminosity.

\begin{figure} 
   \centering
   \includegraphics[width=0.45\textwidth]{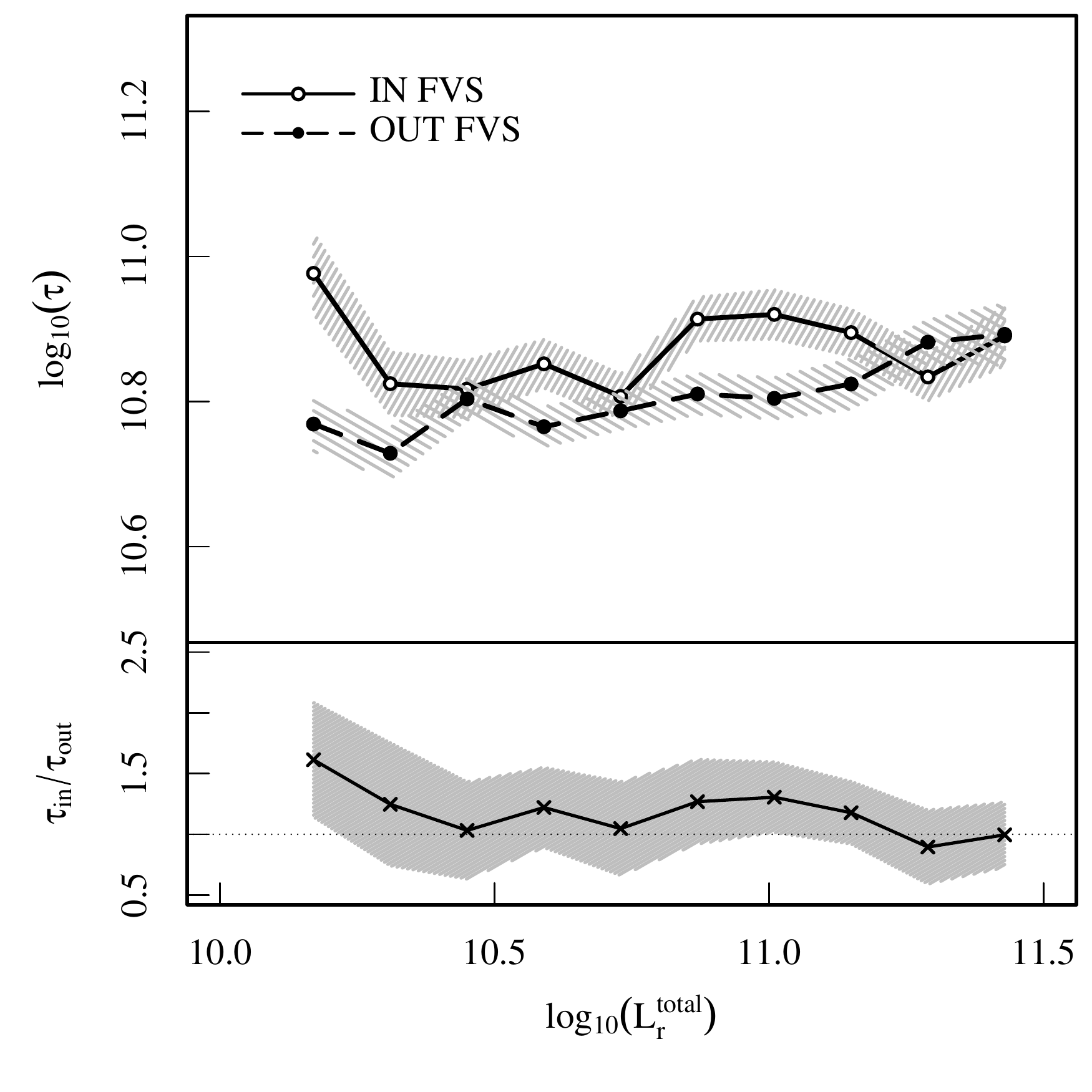}
   \caption{Upper panel: Mean stellar time--scale of groups, in bins of group mean
       luminosities, for groups in $G_{in}$ (solid line) and $G_{out}$ (dashed line).
        The error bands correspond to the error of the means.
        Bottom panel: ratio of stellar time--scale for group samples 
       $G_{in}$ and $G_{out}$.
   }
   \label{F_6}
\end{figure}

With the aim of assigning an indicator of the stellar time-scale at the 
present star--formation rate for groups we use a parameter defined
as $\tau=$M$^{*}/$SFRT, analogous to the one defined for galaxies. 
Thus, $\tau$ provides an estimate of the time--scale for the formation of the total stellar 
mass of the group at the present rate of star formation.
Figure \ref{F_6} displays the results for groups of samples $G_{in}$ and 
$G_{out}$ showing a systematic trend for larger $\tau $ values for groups in FVS. 
It can also be seen that this tendency is larger at lower group luminosities.

\begin{figure}
   \centering
   \includegraphics[width=0.45\textwidth]{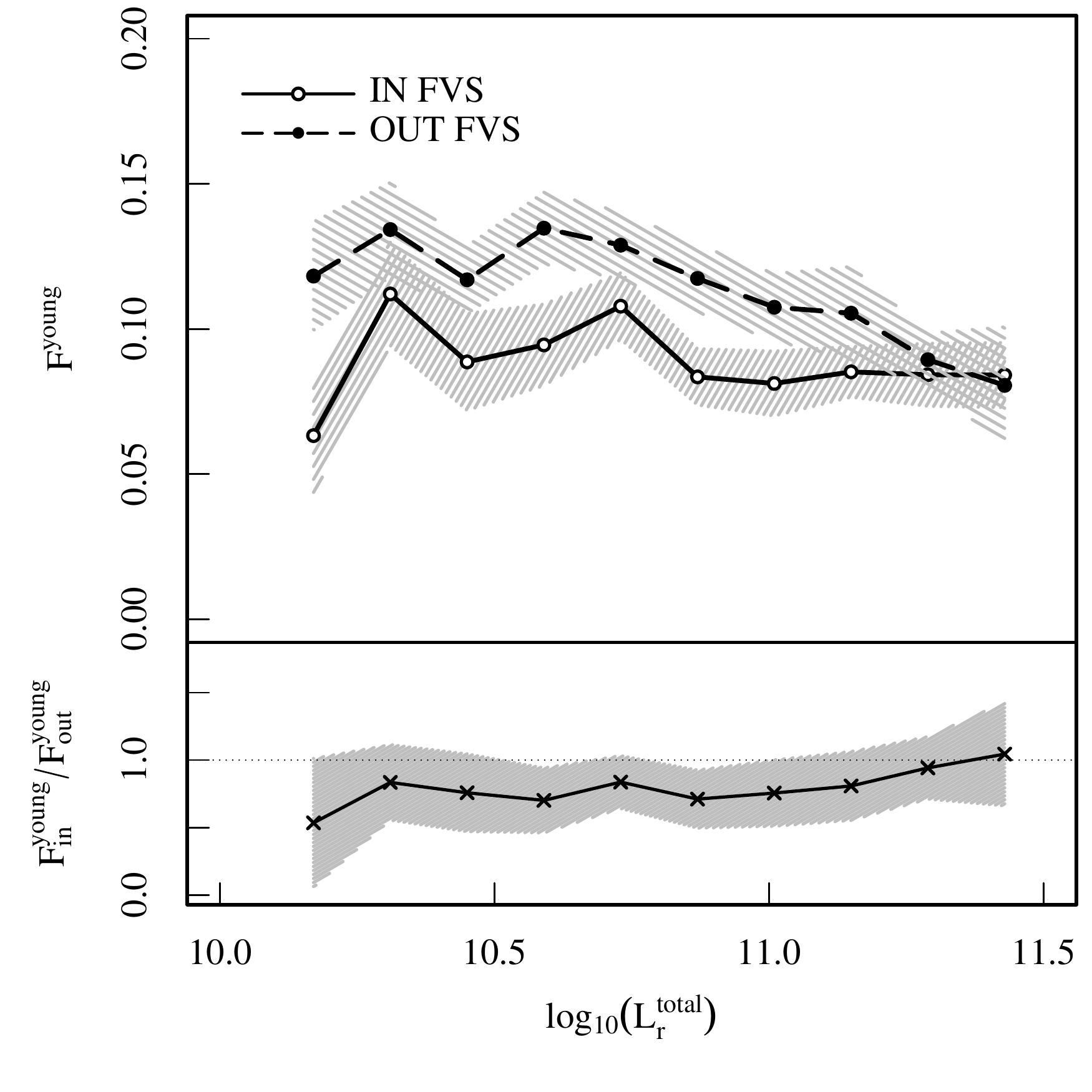}
   \caption{Fraction of young ($Dn_{4000}<1.5$) galaxies in group samples
        $G_{in}$ and $G_{out}$ ($F^{young}$).The error bands correspond to the error of the means.
        Bottom panel: ratio of $F^{young}$ for group samples $G_{in}$ and $G_{out}$.
   }
   \label{F_7}
\end{figure}

We have computed the fraction of galaxies with spectra dominated by a young stellar 
population as revealed by the $Dn_{4000}$ parameter.
Figure \ref{F_7} shows the mean values of $F^{young}$, such that $Dn_{4000}<1.5$, for 
groups in samples $G_{in}$ and $G_{out}$. 
As it can be seen, galaxies dominated by a young stellar population are more frequently found 
in groups not belonging to superstructures. 
We also find that this effect is stronger at lower group luminosities.

\subsection{Clustering properties}

The aim of this section is to study the clustering of galaxies around
group centres for $G_{in}$ and $G_{out}$ samples.
For this purpose, we use two--point correlation statistics following
the procedures described in Section 3 of \citet{Yaryura:2012}.
The two--point galaxy--group 
cross--correlation function, $\xi (r)$, is defined as
the excess of the probability of finding a galaxy, 
at a given distance from a group centre.
Since we use redshift--space positions of galaxies, 
we estimate the correlation function \mbox{$\xi$}, as a
function of the projected ($\sigma$) and line of sight ($\pi$)
distances.
Then, we implement the inversion method presented by \citet{Saunders:1992} to
obtain the spatial correlation function $\xi(r)$ from
\mbox{$\xi(\sigma,\pi)$}.
We integrate along the line of sight to obtain the projected
correlation function $\Xi(\sigma)$:

\begin{equation}
\Xi(\sigma) = 2 \int^{\infty}_{0} \xi(\sigma, \pi) d\pi = 2 \int^{\infty}_{0} \xi(\sqrt{\sigma^{2} + y^{2}}) dy .
\label{eq:xi_sigma_int}
\end{equation} 
\noindent

We estimate the real space correlation function by the
inversion of $\Xi(\sigma)$ assuming a step function 
\mbox{$\Xi(\sigma_i)= \Xi_{i}$} in bins centered in $\sigma_{i}$ and interpolating
between $r = \sigma_{i}$ values \citep[equation 26
of][]{Saunders:1992}:

\begin{equation}
\xi(r) = - \frac{1}{\pi} \sum_{j \geq i} \frac{\Xi_{j + 1} - \Xi_{j}}{ \sigma_{j + 1} -\sigma{j}} ln \left( \frac{ \sigma_{j + 1} + \sqrt{\sigma_{j + 1}^{2} - \sigma_{i}^2}}{ \sigma_{j} + \sqrt{\sigma_{j}^{2} - \sigma_{i}^2}} \right) .
\label{eq:xi_sigma_i}
\end{equation}

\begin{figure} 
  \centering
  \includegraphics[width=0.45\textwidth]{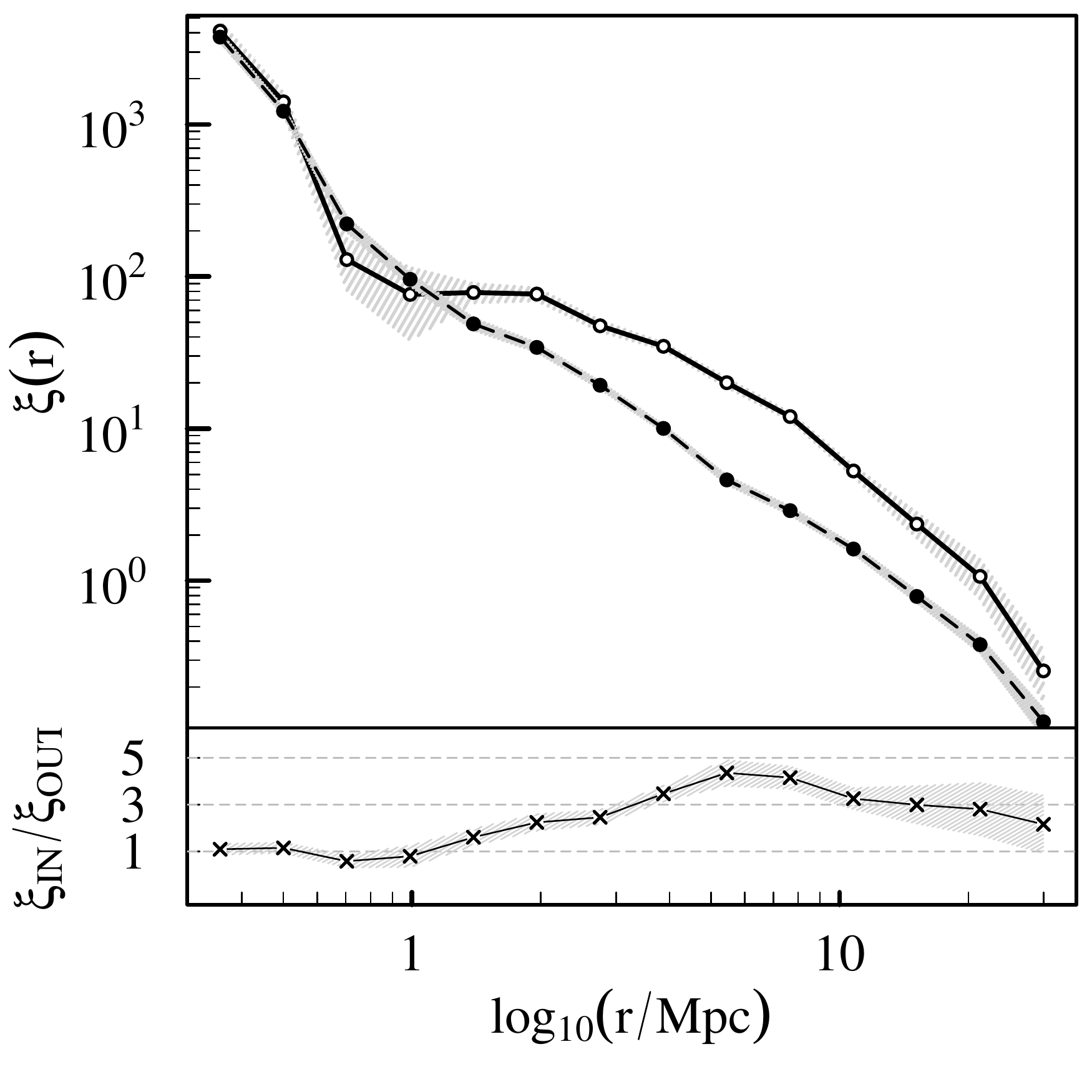}
  \caption{Group--galaxy cross--correlation functions for samples $G_{in}$ (solid line)
      and $G_{out}$ (dashed line), where the groups have the same total luminosity
      distributions. Shaded errors corresponds to  Jacknife uncertanties. 
   }
   \label{F_corr}
\end{figure}

\noindent
In terms of the halo model (\citet{Cooray:2002} and references therein) all
galaxies are associated with haloes, which are defined as dense objects that
constitute the non--linear density field.
In this context, the two--point correlation function can be interpreted as the sum
of two types of galaxy pairs: pairs in the same halo (1--halo term), and pairs in separated
 haloes (2--halo term).
On small scales ($r \lesssim 1 Mpc$) the 1--halo terms dominates, while for larger
scales it becomes negligible predominating the 2--halo term \citep{Sheth:2005,Zehavi:2004}. 
We consider galaxies with r--band luminosities \mbox{$M_{r} < -20.5$}
and take the geometrical centres of groups as centres for the
cross--correlation calculation.
Figure \ref{F_corr} shows the resulting cross--correlation functions
for group centres of samples $G_{in}$ and $G_{out}$.
The solid (dashed) lines correspond to sample $G_{in}$, ($G_{out}$).
As it can be seen in this figure, there are no significant differences
in the 1--halo term between the two correlation functions. 
Thus, the galaxy density profile of groups residing in FVS and
elsewhere are remarkably similar.
These results are consistent with \citet{Yaryura:2012} who analysed
the 2--halo term difference between groups residing in and out FVS.

\subsection{Estimates of cluster sizes} \label{S_rad}

It is difficult to address a reliable characteristic size of a 
galaxy group in observational data. 
The virial radius $r_V$, is often used to represent groups spatial extent.
By definition, and assuming galaxy masses to be known, the virial
radius is related to the total energy of the system through the Virial
Theorem.
The resulting virial radius estimate is sensitive to small separations, leading in
some cases to a significant overestimation of group size.

The minimal enclosing circle is a suitable alternative estimate of group size 
which gives a representative measure of the group boundaries.
In two dimensions, the general minimal enclosing circle problem consists of 
searching the smallest possible circle that encloses all the members 
\citep{sylvester_question_1857}.
Since it is a classical problem of computational geometry, several
algorithms have been developed to solve it efficiently 
(e.g. \citet{welzl_smallest_1991,
gartner_fast_1999,Mordukhovich_smallest_2011}).
The minimal enclosing circle has been used to define the
characteristic size of galaxy systems, in particular in studies of 
compact groups \citep{hunsberger_luminosity_1998}.

\begin{figure}
   \centering
   \includegraphics[width=0.45\textwidth]{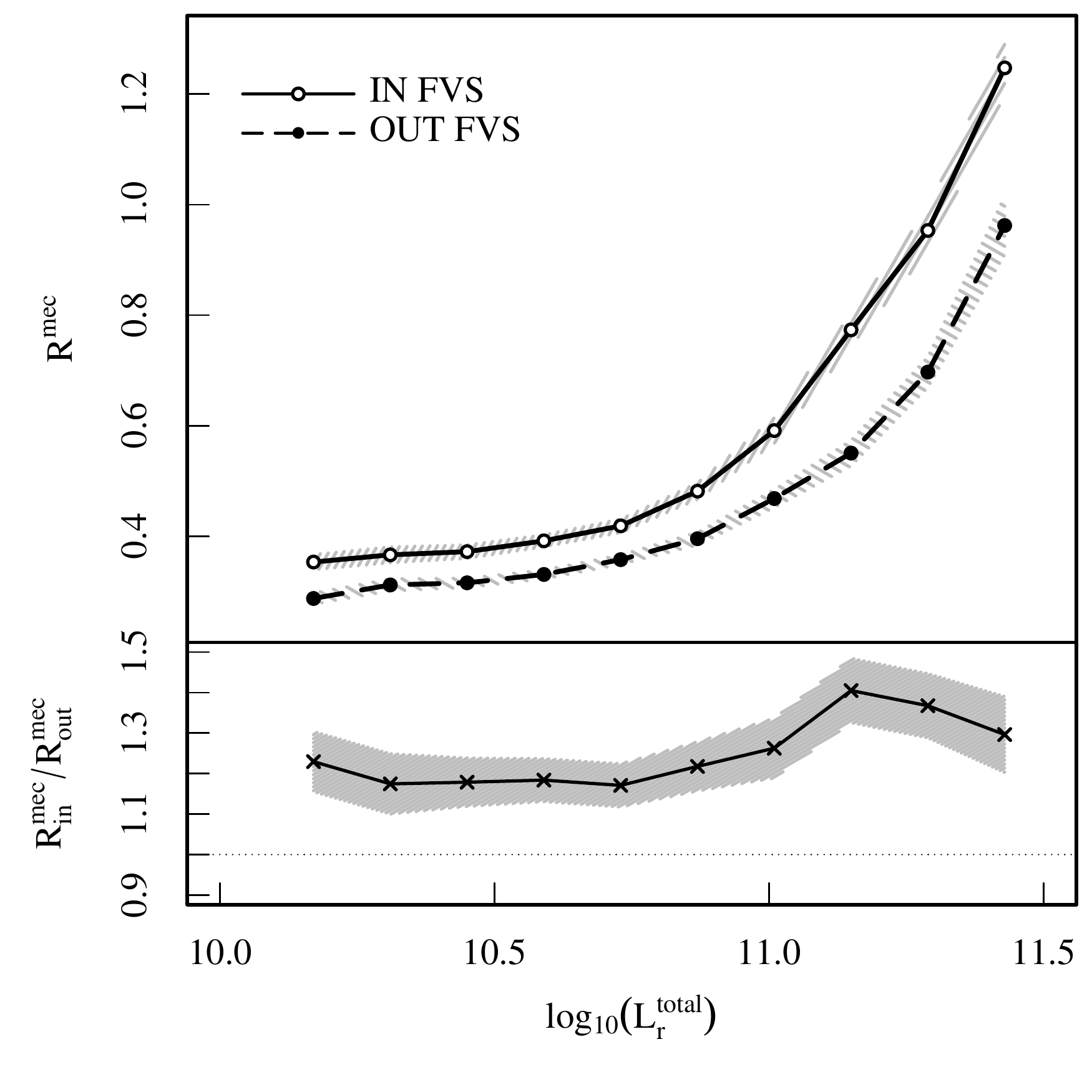}
   \caption{Upper panel: Mean radius of the minimal enclosing circle of groups, in bins of group mean
       luminosities, for groups in $G_{in}$ (solid line) and $G_{out}$ (dashed line).
        The error bands correspond to the error of the means.
        Bottom panel: ratio of the radius of the minimal enclosing circle for group samples 
       $G_{in}$ and $G_{out}$.
   }
   \label{F7}
\end{figure}

We have computed the minimal enclosing circle radii ($R^{mec}$)
for the groups of samples $G_{in}$ and $G_{out}$ using the 
\textsc{miniball} software \citep{gartner_fast_1999}. 
The results are shown in figure \ref{F7}.
According to this figure, at a given total group luminosity, 
group sizes (as estimated via $R^{mec}$) residing in FVS are larger 
than those of groups elsewhere.
We suggest that the FoF group finding algorithm could influence this 
result since the neighbourhood of groups in FVS has a higher background 
level that may link surrounding substructures and induce larger systems.

\subsection{Internal Structure of Groups}

\begin{figure*}
   \centering
   \includegraphics[width=0.98\textwidth]{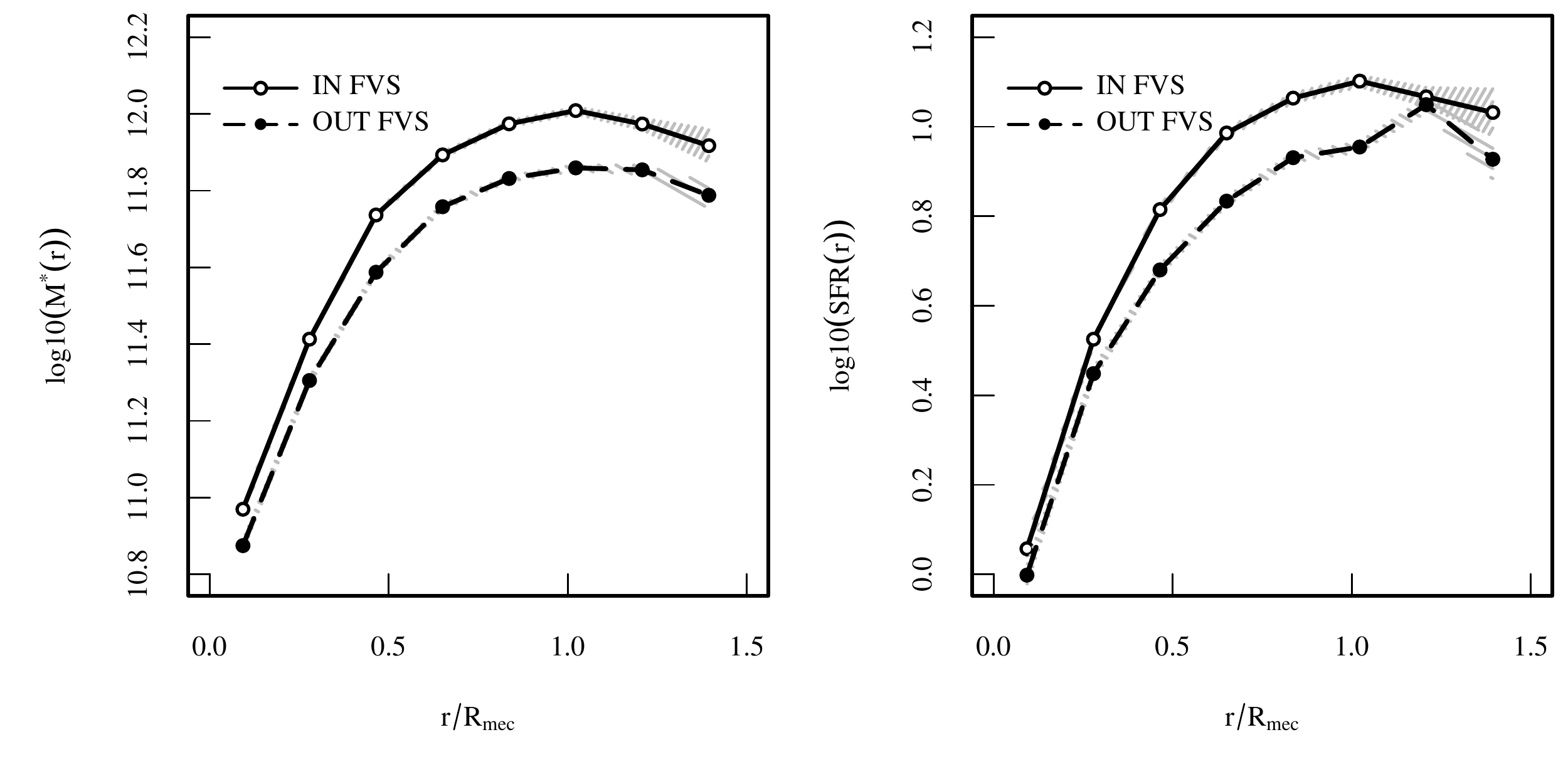}
   \caption{Group radial profiles of stellar mass (left panel) and total star--formation rates
            (right panel) for group samples $G_{in}$ and $G_{out}$.
            The error bands correspond to the error of the means.
   }
   \label{F_8}
\end{figure*}

We have used the minimal enclosing circle radius defined in section \ref{S_rad} 
to normalize the radial distance of galaxies in a group with respect to the geometrical 
center, in order to analyse the stellar mass and star--formation rate profiles.
We have stacked groups of samples $G_{in}$ and $G_{out}$ to compute mean radial profiles
of the stellar mass and star--formation rate (Fig. \ref{F_8}).

\begin{figure}
   \centering
   \includegraphics[width=0.45\textwidth]{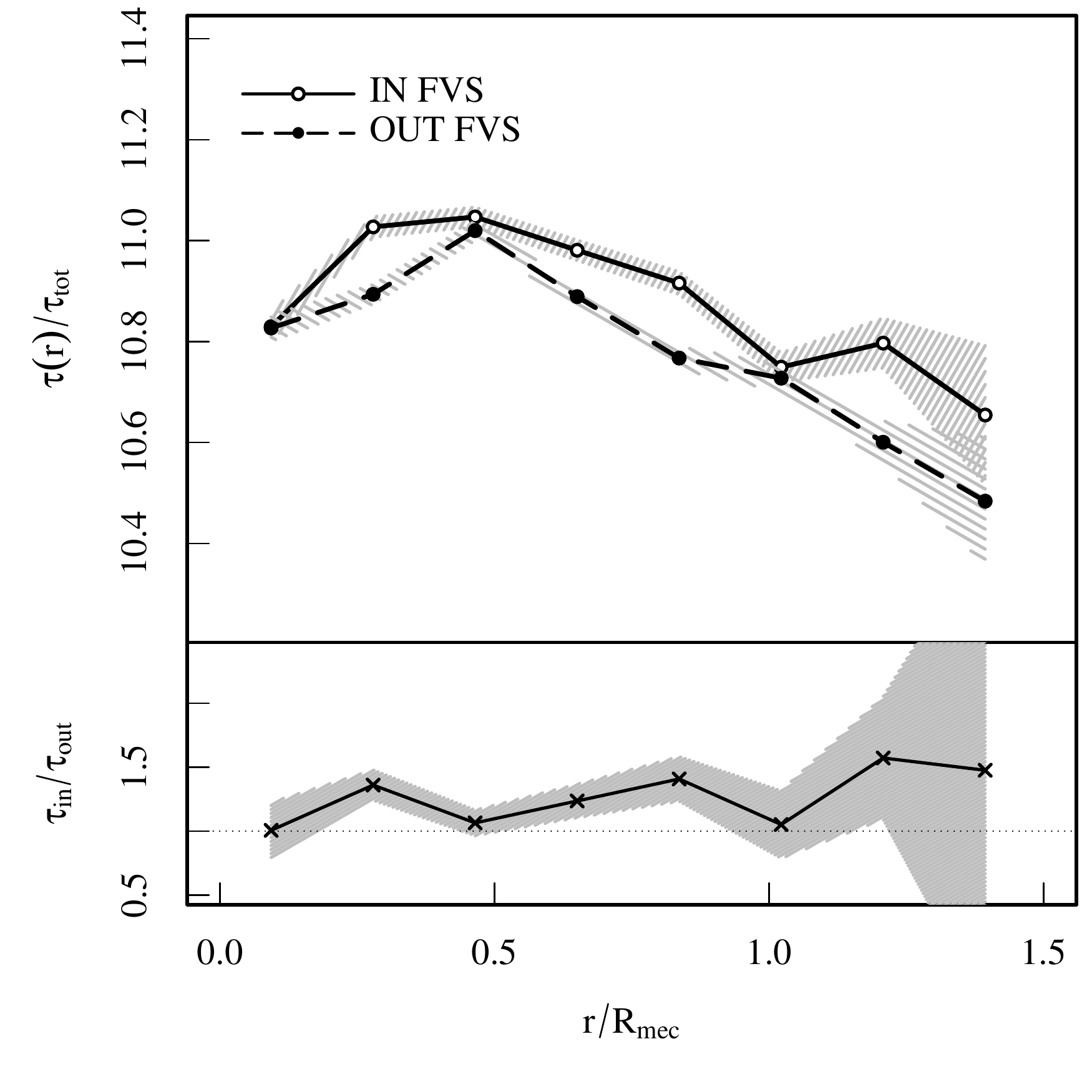}
   \caption{Group radial profile of mean stellar time--scale of group samples
        $G_{in}$ and $G_{out}$.The error bands correspond to the error of the means.
        Bottom panel: ratio of the mean age for group samples 
       $G_{in}$ and $G_{out}$.
   }
   \label{F_9}
\end{figure}

We find that the both, the stellar mass profile and the star--formation rate of  
groups in samples $G_{in}$ and $G_{out}$ differ from each other.
$G_{in}$ systems show a larger stellar mass content and star--formation 
rate, and imply 
a larger time--scale for star formation at all distances from the 
group center.
Figure \ref{F_9} shows the mean $\tau$ parameter as a function of group--centric 
distance for samples
$G_{in}$ and $G_{out}$ showing a similar behaviour althogh $\tau$ values in 
$G_{in}$ are systematically larger than the $G_{out}$ counterpart at the same 
group--centric distance.

\section{Group global properties and their dependence on the global environment} \label{S_discussion}

According to the current modelling of the hierarchical
clustering scenario, the mass of dark matter halos is the main
responsible of their properties \citep{Bond:1991, Lacey:1993,
Mo:1996,  Sheth:2001}.
However, this assumption ignores correlations between different
spatial scales, implying that the large--scale environment where the
halo resides has no influence on its formation and evolution.
\citet{Sheth:2004} find evidence that haloes in dense regions form
earlier than haloes of the same mass embedded in less dense regions.
Using numerical N--body simulations, \citet{Gao:2005} show that halos
assembled at high redshift are more clustered than those of the same
mass that assembled more recently.
This effect, known as ``assembly bias'', stays that the properties of
a given mass halo depend on its formation history.
Hence, if galaxy clusters and groups residing in superstructures have
been evolving under different environmental conditions, it is expected
to find differences in their present day properties.  
In this context, \citet{Einasto:2005} use numerical simulations to
show that the density distribution in large low density regions (such
as voids) present slow evolution, that cease at intermediate
redshifts.
On the other hand, in higher density regions (superclusters) the
structures begin to form early and continue evolving until the
present.
\citet{Wang:2008} find that SDSS galaxy groups with a red central
galaxy are more strongly clustered than groups of the same mass
hosting a blue central galaxy. 
Besides the differences on clustering amplitudes, \citet{Zapata:2009}
observe that galaxy groups in narrow ranges of masses but diverse
formation histories present different galaxy populations.
Analysis of substructure on low density regions manifest that the
luminosity and local density are not completely responsible  of galaxy
properties.
\citet{Ceccarelli:2008} asset that galaxies inhabiting void walls are
systematically bluer and more actively star--forming than field
galaxies at a given luminosity and local galaxy density.
\\ Since the highest density peaks, represented by superstructures,
are expected to be the first sites of gravitational collapse, the
assembly bias scenario can provide some light on our results.
As our group samples have the same luminosity distributions (a proxy
for mass distributions), differences found in their properties can be
then associated to the large--scale environmental effects.
In agreement with \citet{Einasto:2012}, we find that groups residing
in superstructures are richer and present larger velocity dispersions
than those in less--dense global regions.
Regarding to the stellar population, the fraction of young galaxies in
groups tends to be small when the systems are located on
superstructures. 
Also, the time--scale computed from the present--day star--formation
rate indicates that groups inhabiting high density regions assembled
earlier than groups elsewhere. 
This is consistent with the previous analysis of \citet{Lietzen:2012}.
It is expected that groups in superstructures would be the first sites
of star formation in the universe and thus are likely to lack a gas
reservoir suitable for present--day star formation activity.

\section{Conclusions} \label{S_conclusus} 
%{{{S_conclusus/*

We have adopted global luminosity, i.e. the sum of group member 
luminosities, as a group parameter.
Using this parameter, differences in properties of groups residing 
in superstructures can be confronted to those elsewhere.
The use of global luminosity in our work is reinforced by the 
fact that numerical simulations show it as a suitable proxy to 
group total mass \citep{Eke:2004a}.
The main results of this work are summarized as follows:

\begin {itemize}

\item We find that groups residing in superstructures have a 
systematically larger stellar mass content exceeding by a factor 
1.3 that corresponding to groups elsewhere.
This relative excess of the stellar mass content is independent 
of group luminosity. 

\item The mean velocity dispersion of galaxies in groups 
residing in superstructures are $\sim 35$ per cent higher than their 
field counterpart.
%m
The minimal enclosing radii of groups in superstructures are 
systematically larger by $20$ per cent. 

\item A time--scale for star formation defined from the present--day 
star--formation rate shows that groups in superstructures are systematically older. 
Consistently, the fraction of galaxies dominated by a young stellar population, 
$Dn_{4000} < 1.5$, shows the same tendency.

\end {itemize} 

The similarity of the star--formation rate and stellar mass group--centric 
radial profiles also reinforce our intrepretation in terms of the assembly 
bias scenario since the differences found are global. 
Therefore, our results provide evidence that groups in superstructures formed 
earlier than elswhere, as expected in the assembly bias scenario.

%}}}S_conclusus/*

%··········································································

\section*{acknowledgements}
%{{{/*
We thank the referee, Heidi Lietzen, for her through review and
highly appreciate the comments and suggestions, which greatly improved
this work.
This work was partially supported by the
Consejo Nacional de Investigaciones Cient\'{\i}ficas y T\'ecnicas
(CONICET), and the Secretar\'{\i}a de Ciencia y Tecnolog\'{\i}a,
Universidad Nacional de C\'ordoba, Argentina.
Funding for the SDSS and SDSS--II has been provided by the Alfred P.
Sloan Foundation, the Participating Institutions, the National Science
Foundation, the U.S. Department of Energy, the National Aeronautics
and Space Administration, the Japanese Monbukagakusho, the Max Planck
Society, and the Higher Education Funding Council for England. The
SDSS Web Site is http://www.sdss.org/.
The SDSS is managed by the Astrophysical Research Consortium for the
Participating Institutions. The of the Royal Astronomical Society
Participating Institutions are the American Museum of Natural History,
Astrophysical Institute Potsdam, University of Basel, University of
Cambridge, Case Western Reserve University, University of Chicago,
Drexel University, Fermilab, the Institute for Advanced Study, the
Japan Participation Group, Johns Hopkins University, the Joint
Institute for Nuclear Astrophysics, the Kavli Institute for Particle
Astrophysics and Cosmology, the Korean Scientist Group, the Chinese
Academy of Sciences (LAMOST), Los Alamos National Laboratory, the
Max-Planck-Institute for Astronomy (MPIA), the Max-Planck-Institute
for Astrophysics (MPA), New Mexico State University, Ohio State
University, University of Pittsburgh, University of Portsmouth,
Princeton University, the United States Naval Observatory, and the
University of Washington.   
The Millenium Run simulation used in this paper was carried out by the
Virgo Supercomputing Consortium at the Computer Centre of the
Max--Planck Society in Garching.  The semi--analytic galaxy catalogue
is publicly available at
http://www.mpa-garching.mpg.de/galform/agnpaper.
Plots are made using R software.
%}}}/*

\footnotesize{
  \bibliographystyle{mn2e}
%  \bibliography{Bibliography.bib}
}

\end{document}